\documentclass[pre,twocolumn,floatfix,aps,showpacs]{revtex4-1}
\usepackage{graphicx}
\usepackage{amssymb}
\usepackage{amsmath}
\usepackage{epsf}
\usepackage{epsfig}
\usepackage{slashbox}
\usepackage{amsfonts}
\usepackage{bm}
\usepackage{ulem}
\usepackage{psfrag}

\long\def\@makecaption#1#2{%
  \vskip\abovecaptionskip
  \sbox\@tempboxa{#1: #2}%
  \ifdim \wd\@tempboxa >\hsize
    #1: #2\par
  \else
    \global \@minipagefalse
    \hb@xt@\hsize{\box\@tempboxa\hfil}%
  \fi
  \vskip\belowcaptionskip}
\makeatother

\begin{document}

\title{Weak disorder: anomalous transport and diffusion are normal yet again}

\author{M. Khoury$^{(1)}$, A. M. Lacasta$^{(2)}$, J. M. Sancho$^{(1)}$,
and Katja Lindenberg$^{(3)}$}

\affiliation{ 
$^{(1)}$
Departament d'Estructura i Constituents de la Mat\`eria, Diagonal 647, E-08028 Barcelona, Spain\\
$^{(2)}$
Departament de F\'{\i}sica Aplicada, Universitat Polit\`{e}cnica de Catalunya,
Avinguda Doctor Mara\~{n}on 44, E-08028 Barcelona, Spain\\
$^{(3)}$
Department of Chemistry and Biochemistry 0340 and BioCircuits Institute, 
University of California, San Diego, La Jolla, California 92093-0340,
USA\\
}

\begin{abstract}
We have carried out a detailed study of the motion of particles driven
by a constant external force over a landscape consisting of a
periodic potential corrugated by a small amount of spatial disorder. 
We observe anomalous behavior in the form of subdiffusion and superdiffusion
and even subtransport over very long time scales. 
Recent studies of transport over slightly random landscapes have focused only
on parameters leading to normal
behavior, and while enhanced diffusion has been identified when the external
force approaches the critical value associated with the transition from
locked to running solutions, the regime of anomalous
behavior had not been recognized. We provide a qualitative explanation for the
origin of these anomalies.
\end{abstract}

\maketitle

Solid state surfaces frequently present periodic potentials marred by some 
disorder. Herein we show that an overdamped particle moving over such
a potential in one dimension (1D) may exhibit anomalous behavior in the form of 
superdiffusion, subdiffusion, and even subtransport. Although we cannot
prove that these are steady state regimes, 
our numerical simulation data show them to be present over
time spans of several orders of magnitude. 

That diffusion of particles over both periodic and random surfaces 
lead to some forms of anomalous behavior is
of course well known and continues to attract a great deal of attention both
theoretically and experimentally~\cite{Theory,anomalous,PhysRevE70,SPIE,katja2,Katja,khoury:diffusion,Lutz,Grier,Reimann}.
In periodic potentials with low friction, extremely long (in time) dispersionless transport
regimes can be observed when forces exceed
a critical force~\cite{Katja}.  Moreover, in these same systems, in both
overdamped and underdamped regimes, the diffusion 
coefficient versus the applied force presents a pronounced peak
around the critical force that allows the coexistence of locked and running
states~\cite{Theory,SPIE,katja2,Katja,khoury:diffusion}. The
The enhancement is quantitatively larger than the free particle
diffusion coefficient.
This behavior has been observed experimentally when
tracking the motion of colloidal spheres through a periodic potential
created with optical vortex traps~\cite{Grier}.

The enhancement of the diffusion coefficient is even more pronounced when
disorder is also present~\cite{Grier}. This phenomenon has been tested by
numerical simulations on a surface in which a small amount of
spatial disorder in the form of a random potential is added to the
periodic potential~\cite{Reimann}.  Dramatic
diffusive enhancement occurs even for very small amounts of
disorder, e.g., when the amplitude of the random contribution of the potential
is as small as $ \sim 5 \%$ of that of the periodic
contribution. 

Although dramatic, diffusive enhancement turns out to be only a limited aspect
of the story because it is not the only manifestation of
disorder. Here we present
a range of additional anomalous transport and diffusion phenomena arising
from weak disorder that have not been previously noted.
Our model and the behaviors it exhibits are inspired by~\cite{Grier,Reimann}.

We consider the overdamped motion of identical noninteracting Brownian
particles moving in a 1D potential landscape $U(x)$ following the Langevin equation
$\gamma \dot{x}(t) = -U'(x) + F + \xi(t)$.
Here $x$ is the position of the particle, $t$ denotes time, $\gamma$ is the
dissipation parameter, $F$ is the applied force, and $\xi(t)$ is 
Gaussian thermal noise at temperature $T$. The correlation function
of the noise obeys the fluctuation-dissipation relation
$\langle \xi(t)\, \xi(t')\rangle = 2 \gamma k_BT\delta(t-t')$.
The potential $U(x)$ consists of a periodic part,
$V_p(x)=V_0\,\cos\left(2\pi x/\lambda_p\right)$,
and a Gaussian spatially random contribution $V_r(x)$ with
correlation function
\begin{equation}
g_r(x)\equiv \langle V_r(x)\,V_r(0)\rangle =
\frac {V_0^2} {2}\exp\left(-\frac {2\pi^2 x^2} {l_r^2}\right).
\label{correlran}
\end{equation}
We need to choose parameters to capture the relative contribution of each, as well as
their relative amplitudes and length scales.  The relative contributions are
determined by the parameter $\sigma$ in the combination
\begin{equation}
U(x)=(1-\sigma) V_p(x)+\sigma V_r(x),\qquad \sigma\in [0,1].
\label{totalpot}
\end{equation}
Furthermore, we have chosen the potential correlations $g_r(x)$ and
 $g_p(x) = (V_0^2/2)\cos\left(2\pi x/\lambda_p\right)$
to be equal at $x=0$,
$g_p(0) = g_r(0)$. This ensures that
the total potential amplitude is of order $V_0$ independently of $\sigma$.
Also, with the particular choice $l_r=\lambda_p$ 
the two correlation functions are identical up to second
order in a Taylor expansion.  

The equation of motion can be rescaled into dimensionless form
in terms of the spatial variable $z=2\pi x/\lambda_p$ and the temporal variable
$\tau=[(2\pi)^2 V_0/\gamma \lambda_p^2]t$.
This yields
\begin{equation}
 \dot{z}=(1-\sigma) f_p(z) + (\sigma/ \lambda) f_r(z/\lambda) + {\mathcal F} + \eta(\tau),
\label{2r}
\end{equation}
where $f_p$ and $f_r$ are the dimensionless forces arising from $V_p$ and $V_r$, respectively, and $\eta$ is the dimensionless noise.
The dimensionless parameters are
\begin{equation}
\lambda= \frac{l_r}{\lambda_p};\qquad {\mathcal F}= \frac{\lambda_p F}{2\pi V_0}; \qquad
{\mathcal T}= \frac{ḱk_B T}{V_0}.
\label{scales}
\end{equation}
Throughout this work we set ${\mathcal T}=0.01$ and $\lambda_p=2\pi$.
Variations in ${\mathcal T}$ do not lead to any additional phenomenology. The
specific choice of $\lambda_p$ is only important for passage between
dimensionless and dimensioned units. An important
note is that in these variables a decrease of $\lambda$ even for fixed $\sigma$
leads to an
increase in the relative contribution of the random force.

\psfrag {F} {$ {\mathcal F}$}

\begin{figure}[h]
\centering
\epsfig{figure=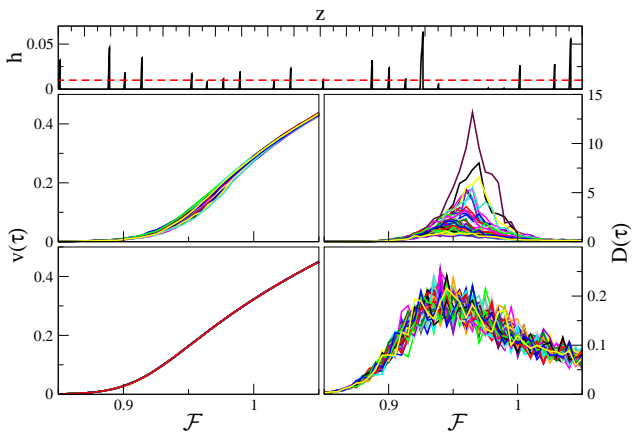,width=7.0cm,angle=-90}
\caption{(Color online) Upper panel: Barriers $h$ of one realization of the total potential
in a spatial domain of $\sim 32$ periods with $\lambda=0.5$ and ${\mathcal F}=1$. Middle and lower panels: Numerical data for 50 different disordered
potentials and $N=100$ particles in each potential. Left panels: $v(\tau)$.
Right panels: $D(\tau)$. Middle panels: $\lambda=1$.
Lower panels: $\lambda= 2\pi$ (regime of Ref.~\cite{Reimann}).
$\sigma=0.05$, $\tau=10^4$.
}
\label{alpha0p95}
\end{figure}

We have carried out numerical simulations of Eq.~(\ref{2r}) over a large number (100) of particle trajectories and a large number of 
realizations of the random potential contribution (typically 50-100) in order to calculate the velocity and diffusion coefficient 
using the  prescriptions
\begin{equation}
v(\tau) = \frac{\langle z(\tau)\rangle}{\tau}, \qquad D(\tau)= \frac{\langle z^2(\tau)\rangle
-\langle z(\tau) \rangle^2}{2\tau}.
\end{equation}
The brackets $\langle \cdots \rangle$ denote averages over many
trajectories and potentials. The numerical procedures are entirely
standard. The usual assumption is that stationary values are reached at long
times, an assumption that we show here to be at the very least
questionable. Representative outcomes of this procedure are shown in the middle
and
lower panels of Fig.~\ref{alpha0p95}. These outcomes qualitatively capture an
unanticipated range of behaviors in the right middle panel.

The velocity reaches a well-defined stationary value for
any value of the force in the figure (left panels), 
even in the regime of sharp increase around ${\mathcal F}=1$. Also,
as noted earlier, for the parameters used in Ref.~\cite{Reimann},
we see the enhancement of the (well defined) diffusion coefficient around 
the critical deterministic
force ${{\mathcal F}}_c$, as discussed in that work (lower right panel). 
However, entirely different behavior is now seen in the middle right panel of the
figure, which shows huge variations in the diffusion coefficient.  Note that these variations extend over orders
of magnitude. The question then is - what leads to these variations and why
were they not identified in Ref.~\cite{Reimann}?

The answer lies in the choice of the correlation length of the random
potential.  In~\cite{Reimann} the correlation length is much greater than a
single period of the periodic potential, that is, the randomness is
very smooth.  However, entirely different behavior is seen when the random
potential is more corrugated, as it is in our case.
We see that the traditional
diffusion coefficient is no longer well-defined in this regime but instead presents
very strong fluctuations around its peak value, making a
precise estimation of $D$ questionable.  Note also the very large scale differences
of the middle and lower right panels of the figure. These are signatures of
anomalous behavior.
While it is not clear whether a well-defined value of $D$ would be
obtained at much longer times beyond our computational reach, we have
repeated these calculations for many more particles and potential realizations
and continue to find this variability. 

To understand the origin of these anomalies, in the upper panel of Fig.~\ref{alpha0p95} we plot the barriers $h$
of the total potential $U(z)-{\mathcal F}z$ for a particular realization of 
the random potential. The heights and locations
of the barriers are
random, and most of them exceed the thermal energy (dashed line). Smaller values of $\lambda$
lead to a greater number and height of the barriers. As a particle moves along such a landscape, it
must overcome these barriers, some of which are extremely high. Indeed, 
$V_r(z)$ allows for barriers of any height, limited only by finite system sizes
and simulation times.
Thus, even with a small amount of disorder the particle motion is dominated by
random waiting times due to the dispersion of the barrier heights, and a few
long waiting times that greatly influence the outcome.
We go on to show that the diffusion anomaly in Fig.~\ref{alpha0p95} is a consequence of
such landscapes and that it is qualitatively different from a simple large enhancement of the
diffusion coefficient. In fact,
strong fluctuations in the usual ensemble calculation
of the diffusion coefficient point to the fact that our system is exhibiting
behavior reminiscent of aging or of
weak ergodicity breaking~\cite{Sokolov,Bouchaud,Klages} over the time scales of our
simulations. We approach the problem with this observation in mind.

Our numerical results are collected as follows. Particles are initially located
at random positions uniformly extended over a region of about 1000
sites, and for each we observe the times $t_p$ that it takes a single
particle to cover the underlying spatial period $\lambda_p$ over the course of its trajectory over
a long time.  We collect these statistics for many particles and many realizations of
the random potential.  If the motion of the particles is ``normal'' then
following the reasoning in~\cite{Theory,Lutz} and also used in~\cite{Reimann}
leads to the average velocity and diffusion
coefficient expressions
\begin{equation}
 v=\frac{\lambda_p}{\langle t_p\rangle},\qquad
D= \frac{\lambda_p^2}{2}\,\frac{\langle  \Delta^2 t_p\rangle }{\langle t_p\rangle^3}.
\label{D}
\end{equation} 
Here $\langle t_p\rangle$ is the average of the $t_p$ over all trajectories, particles, and potential 
realizations and $\langle  \Delta^2 t_p\rangle$ is their dispersion about the average. 
It is of course evident that these expressions are only well defined if the first and
second moments of the distribution of the $t_p$ are finite.
We present numerical evidence that
indicates that in the presence of a small amount of disorder in the potential 
the distribution $P(t_p)$ of the time $t_p$ to cover a single period
can have a power-law tail for long stretches of time,
\begin{equation}
P(t_p) \sim t_p^{-\beta}.
\label{tdistribution}
\end{equation}
This behavior leads to the
observed anomalies if $\beta$ is sufficiently small.
For $\beta > 3$, the first and second moments are finite, so transport and
diffusion are normal. 
In the range $2<\beta<3$, the first moment is finite but the second moment diverges. This leads
to a finite average velocity $v$, but the
diffusion coefficient as defined in Eq.~(\ref{D}) diverges (superdiffusion). In the
interval $3/2 < \beta < 2$ the first moment diverges
and the average velocity thus vanishes (subtransport). 
The second moment again diverges (superdiffusion). For $1 < \beta < 3/2$ we
again have substransport, and the diffusion coefficient decays to zero (subdiffusion).
In this case the particle remains extremely localized around its point of
origin. Interestingly, we observe all of these behaviors.

\begin{figure}[h]
\centering
\epsfig{figure=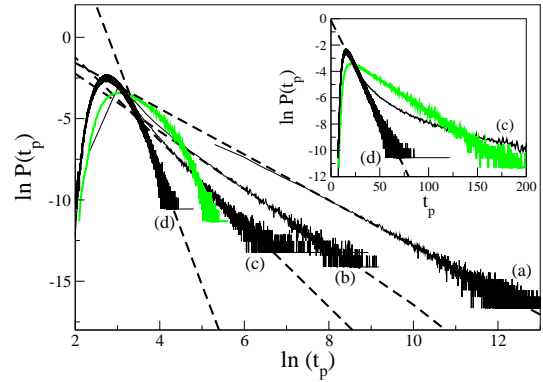,angle=-90,width=8cm}
\caption{(Color online) Log-log plot of the time distribution for different
values of $\lambda$. (a) $\lambda=0.1$ ($\beta=1.4$), (b) $\lambda=0.4$ ($\beta=
1.8$), (c) $\lambda=0.8$ ($\beta = 2.6$), and (d) $\lambda=2\pi$ ($\beta =6.9$).
The parameter values are ${\mathcal F}=1$ and $\sigma=0.05$.
Inset: Log plot of $P(t_p)$ for cases (c) and (d).
Also shown is the
exponential distribution for a purely periodic potential
with the same force (green online).Time $\tau\ge 10^4$ as needed to obtain
reliable histograms.}
\label{time}
\end{figure}

To obtain values of $\beta$, we focus on the tails of the distribution.
If they follow a power law, exponents $\beta$ are estimated,
as depicted in Fig.~\ref{time}.
In the upper panel of Table~\ref{table}, we present a set of typical results of simulations
leading to different values of the exponent $\beta$, obtained for two levels of
disorder and
for different forces.
We observe that, in many cases, a value $\beta < 3 $ 
associated with anomalous behavior is found. Table~\ref{table} also
shows that a lower level of disorder ($\sim 2 \%$) leads to higher exponents.
These estimated values are of course informative but should not be interpreted too narrowly because
they are based on sparse statistics.
For the sake of comparison with Ref.~\cite{Reimann}, we also show
the case $\lambda=2\pi$ 
for a $5\%$ level of disorder. We observe what we expected from Eq.~(\ref{2r})
when
$\lambda$ increases, that is, the effect of the random part
of the potential 
is greatly muted.  This implies that most of the exponents are in the range
$\beta >3$ (normal behavior) for the range of forces selected, which explains
why the anomalies explored herein were not found in Ref.~\cite{Reimann}. 
Much smaller forces would need to
be explored to find robust anomalous behavior there.  For our parameter choices,
using $\lambda$ as a control parameter (keeping $\mathcal F$ fixed)
also leads to $\beta$ values covering all the possible regimes (see bottom panel of Table~\ref{table}).

\begin{center}
\begin{table}[h]
\resizebox*{8cm}{!}{
\begin{tabular}{c||c|c|c|c|c|c|c|}
\backslashbox{$\sigma$}{${\mathcal F}$} & 0.82 & 0.92 & 0.95 & 0.98 & 1.0 & 1.02 & 1.05\\
\hline \hline
0.05 ($\lambda=1$) & 1.1 & 1.7 & 2.2 & 2.7 & 2.9 & 3.6 & 4.8 \\
\hline
0.02 ($\lambda=1$) & 1.4 & 2.3 & 3.2 & 4.3 & 4.6 & $>5$ & $>5$ \\
\hline \hline
0.05 ($\lambda=2\pi$) & 2.2 & 3.0 & $>5$ & $>5$ & $>5$ & $>5$ & $>5$ \\
\hline 
\multicolumn{8}{c}{ } \\ 
\backslashbox{$\sigma\,$}{$\,\lambda$} &0.1& 0.4 & 0.6 & 0.8 & 1.2 & 1.4 & $2\pi$\\
\hline \hline
0.05 (${\mathcal F}=1$) & 1.4 & 1.8 & 2.1 & 2.6 & 3.5 & 4.0 & $>5$ \\
\hline
\end{tabular}}
\caption{Upper panel: Exponents $\beta$ for different values of the force and two values of $\sigma$ and $\lambda$. Bottom panel: Exponents $\beta$ for 
different values of $\lambda$, with ${\mathcal F}=1$ and $\sigma=0.05$}
\label{table}
\end{table}
\end{center}

As the value of $\beta$ increases with increasing $\lambda$, the shape of the distribution
$P(t_p)$, particularly its tail, changes from a power law form to the more typical
exponential associated with normal behavior. This occurs because changing
$\lambda$ induces a change of the effective contribution of the
random part of the potential, 
as already noted earlier. We pointed out that increasing $\lambda$ diminishes
the effects of the disorder. In fact, case (d) of
Fig.~\ref{time} is no longer a power law, as shown in the inset
of the figure, but is instead better described as an exponential,
characteristic of normal diffusive behavior.

\begin{figure}[h]
\centering
\epsfig{figure=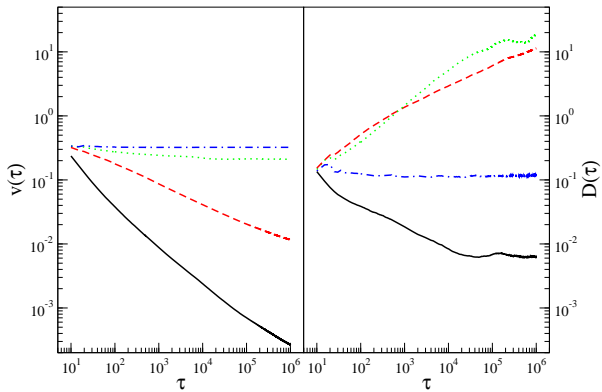,angle=-90,width=8.5cm}
\caption{(Color online) Log-log plot of the time evolution of the velocity (left) and of the
diffusion coefficient (right) for different values of $\lambda$: $\lambda=0.1$ (solid line), 
$\lambda=0.4$ (dashed line), $\lambda=0.8$ (dotted line) and $\lambda=2\pi$ (dotted-dashed line).
Parameter values: $\sigma=0.05$, ${\mathcal F}=1$. Number of potential
realizations: 100, except for $\lambda=2\pi$ where 20 realizations are
sufficient.  
}
\label{trajectories}
\end{figure}
Finally, temporal evolutions of the velocity and the diffusion coefficient for
the four different
regimes of behavior are shown in Fig.~\ref{trajectories}. 
They are in qualitative agreement with this analysis, at 
least during a transient of several decades. The subtransport regimes of the
two cases with the lowest values of $\lambda$ are clearly seen, as are the
subdiffusive and superdiffusive behaviors of the diffusion coefficient.

In conclusion, we have carried out a detailed numerical study of the motion of particles driven by
a constant external force over a one-dimensional landscape consisting of a periodic potential modified
by a small amount of spatial disorder.
We have identified a set of dramatic anomalous behaviors as diverse as
subtransport, subdiffusion, and superdiffusion on the same surfaces as the
driving force or the random corrugation length is varied. These behaviors are
observed over very long time
scales.  Their asymptotic persistence behavior is not known. Earlier studies
have focused only on parameters of normal
behavior~\cite{Grier,Reimann}, and while they have identified
the occurrence of enhanced diffusion when the external force
approaches the critical value associated with the transition from
locked to running solutions, they have not recognized the regime of anomalous behavior.

The regimes that exhibit these anomalous behaviors are identified by the correlation length $l_r$
of the random portion of the potential.  We find anomalous behavior when $l_r\lesssim \lambda_p$, the
period of the periodic portion, that is, when the periodic potential is
slightly corrugated over short distances. 
Earlier studies had focused on the regime $l_r = 2\pi \lambda_p$, that is, on
regimes of very long smooth variation of the random contribution to the
potential.
Experiments with more corrugated surfaces than have been used so far~\cite{Grier} are
clearly desirable to see the effects that we have identified.

It would of course also be desirable to extend these studies to higher
dimensions and to find an analytic characterization of these systems, one which
would allow insight into the asymptotic behavior. Such an analysis has recently been presented for
the case of a piecewise linear random potential~\cite{Denisov}, but seems not yet to be available
for the more realistic potentials considered here.  


This work was supported by the MICINN (Spain) under the project
FIS2009-13360 (AML, MK and JMS), by Generalitat de Catalunya
Projects 2009SGR14 (JMS, MK) and 2009SGR878 (AML), and the grant FPU-AP2005-4765 (MK).
KL gratefully acknowledges the NSF under Grant No. PHY-0855471.

\end{document}